\documentclass[preprint,superscriptaddress,amsmath,amssymb,prd,aps,showpacs,floatfix,nofootinbib]{revtex4-1}

\usepackage[T1]{fontenc}
\usepackage{lmodern}

\usepackage[dvipsnames,x11names]{xcolor}

\usepackage[colorlinks=true,linkcolor=Blue,citecolor=ForestGreen,urlcolor=NavyBlue]{hyperref}

\usepackage{graphicx}
\usepackage[sort&compress]{natbib}
\usepackage{lipsum}
\usepackage{morefloats}
\usepackage[pdf]{pstricks}
\usepackage{subfigure}
\usepackage{amsmath}
\usepackage{amssymb}
\usepackage{amsfonts}
\usepackage{rotating}
\usepackage{cancel}
\usepackage{mathtools}
\usepackage{bbm}
\usepackage{dsfont}
\usepackage{bbold}
\usepackage{multirow}
\usepackage{ulem}
\usepackage{physics}

\newcommand{\mev}{\textrm{ MeV}}

\newcommand{\GXNU}{\affiliation{Department of Physics, Guangxi Normal University, Guilin 541004, China}}

\newcommand{\GXZD}{\affiliation{Guangxi Key Laboratory of Nuclear Physics and Technology, Guangxi Normal University, Guilin 541004, China}}

\newcommand{\CSU}{\affiliation{School of Physics, Central South University, Changsha 410083, China}}

\newcommand{\ITPCAS}{\affiliation{CAS Key Laboratory of Theoretical Physics, Institute of Theoretical Physics,
Chinese Academy of Sciences, Beijing 100190, China}}

\newcommand{\HZU}{\affiliation{ School of Science, Huzhou University, Huzhou 313000, Zhejiang, China}}

\newcommand{\IFIC}{\affiliation{Departamento de F\'{\i}sica Te\'orica and IFIC, Centro Mixto Universidad de
Valencia-CSIC Institutos de Investigaci\'on de Paterna, Apartado 22085,
46071 Valencia, Spain}}

\begin{document}

\frenchspacing

\title{\boldmath Triangle singularity in the $J/\psi \to \phi \pi^+ a_0^-(\pi^- \eta),\; \phi \pi^- a_0^+(\pi^+ \eta)$ decays}

\author{C. W. Xiao}
\email{xiaochw@gxnu.edu.cn}
\GXNU
\GXZD
\CSU

\author{J. M. Dias}
\email{jorgivan.mdias@itp.ac.cn}
\ITPCAS

\author{L. R. Dai}
\email{dailianrong@zjhu.edu.cn}
\HZU

\author{W. H. Liang}
\email{liangwh@gxnu.edu.cn}
\GXNU
\GXZD

\author{E. Oset}
\email{Oset@ific.uv.es}
\GXNU
\IFIC


\begin{abstract}
  We study the $J/\psi \to \phi \pi^+ a_0(980)^- (a_0^- \to \pi^- \eta)$ decay, evaluating the double mass distribution in terms of the $\pi^- \eta$ and $\pi^+ a^-_0$ invariant masses. We show that the $\pi^- \eta$ mass distribution exhibits the typical cusp structure of the $a_0(980)$ seen in recent high statistics experiments, and the $\pi^+ a^-_0$ spectrum shows clearly a peak around $M_{\rm inv}(\pi^+ a^-_0)=1420 \,{\rm MeV}$, corresponding to a triangle singularity. When integrating over the two invariant masses we find a branching ratio for this decay of the order of $10^{-5}$, which is easily accessible in present laboratories.
  We also call attention to the fact that the signal obtained is compatible with a bump experimentally observed in the $\eta \pi^+\pi^-$ mass distribution in the $J/\psi \to \phi \eta \pi^+\pi^-$ decay and encourage further analysis to extract from there the $\phi \pi^+ a_0^-$ and $\phi \pi^- a_0^+$ decay modes. 
\end{abstract}

\maketitle

\section{Introduction}

Triangle singularities (TS), introduced in Refs.~\cite{karplus,landau}, correspond to processes in which there is a triangle diagram in the amplitude, where the three intermediate particles can be placed simultaneously on shell, representing a reaction that can occur at the classical level \cite{norton}, in which case the amplitude becomes infinite in the limit of zero width of the intermediate particles.
The issue has experienced a revival in recent years, one of the reasons, among many, being the fact that after early claims by the COMPASS Collaboration of the ``$a_1(1420)$'' discovery \cite{adolf}, it was soon explained as a consequence of a triangle singularity in which the $a_1(1260)$ resonance decays into $K^* \bar{K},\, K^* \to \pi K$, and then $K\bar{K}$ fuse to give the $f_0(980)$ resonance, the $\pi f_0(980)$ being the observed decay mode \cite{qzhao,misha,acetidai,Alexeev}. 
Earlier than that, there was an interpretation of the isospin violating decay $\eta(1405) \to f_0(980) \pi^0$ \cite{beseta} also in terms of a triangle singularity \cite{wuzou,fcaliang,wuwu} (see also Refs.~\cite{Acha1,Acha2} showing a reduction of the absolute rate of the reaction when the width of the $K^*$ is considered), 
and more recently the interpretation of the $p p \to \pi^+ d$ reaction \cite{serre,said} in terms of a triangle singularity \cite{ikenomolina}. 
More examples were given in the interpretation of the $J/\psi \to \eta \pi^0 \phi$ reaction in Ref.~\cite{jingsakaiguo}, and in the enhanced isospin violation in $D_s^+ \to \pi^+\pi^0 a_0(980) [f_0(980)]$ or $\bar{B}_s^0 \to J/\psi \pi^0 a_0(980) [f_0(980)]$ \cite{sakailiang,sakailiaxie}. 
A long list of reactions showing effects of triangle singularities can be seen in Table 1 of the review paper \cite{reviewts}. 
Also, a reformulation of the TS has been provided in Ref.~\cite{bayaracetiguo}, which is at the same time pedagogical and practical. 

\begin{figure}[t]
  \centering
  \includegraphics[scale=0.65]{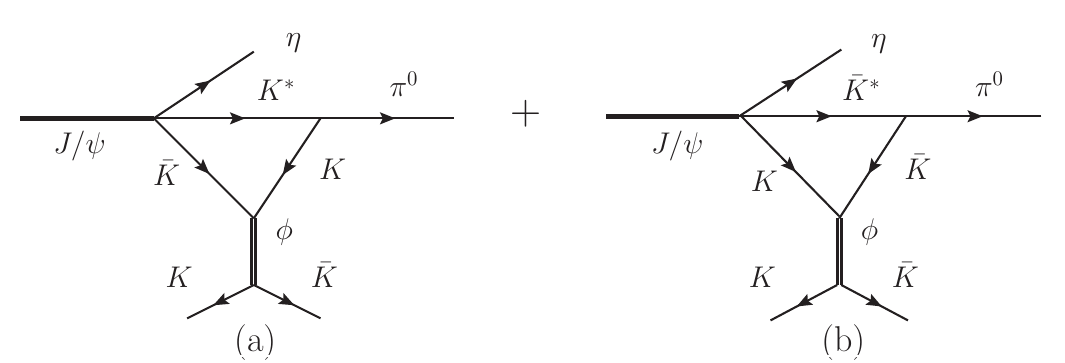}
  \includegraphics[scale=0.68]{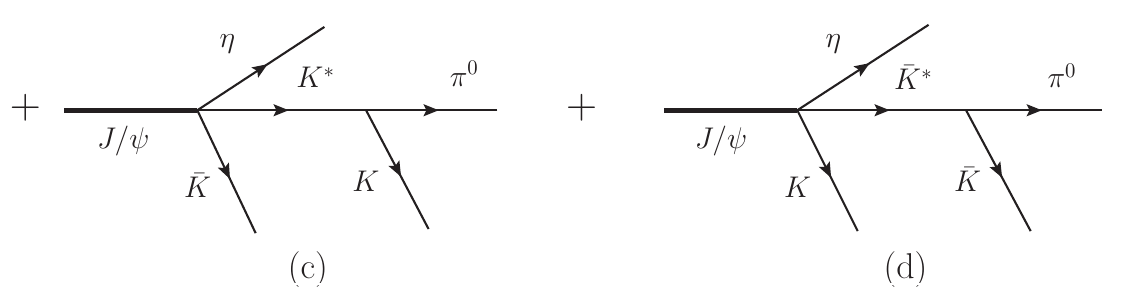}
  \caption{Mechanisms in $J/\psi \to \eta \pi^0 \phi \; (\phi \to K\bar{K})$: (a), (b) TS; (c), (d) the tree level.}
  \label{fig:mech1}
  \end{figure}

The issue has emerged once more due to the recent BESIII paper \cite{besrecent}, improving considerably on earlier measurements of the $J/\psi \to \eta \pi^0 \phi$ reaction, where the ideas of Ref.~\cite{jingsakaiguo} could be tested. 
Indeed, that reaction develops a TS in the $\pi^0 \phi$ mass distribution, peaking around $M(\phi \pi^0) \sim 1400 \mev$, where a clear signal is seen in Ref.~\cite{besrecent}. 
Yet, the interpretation of the peak is delicate, because, as discussed in Ref.~\cite{jingsakaiguo}, the peak is masked by a tree level contribution which also has a singular behavior. 
Indeed, the TS mechanism proceeds via the diagrams of Figs.~\ref{fig:mech1} (a) and (b), but the same final state is reached by the tree level diagrams of Figs.~\ref{fig:mech1} (c) and (d).
Then, Schmid theorem \cite{schmid} comes into play because if the triangle diagrams develop a TS they can be reabsorbed into the tree level diagrams with a simple change of phase, and the decay width is not affected by the TS diagrams. 
A thorough review of this issue was done in Ref.~\cite{vinischmid}, showing that the theorem holds in the limit of the $K^*$ width (in the present case) going to zero, and when there are no inelastic channels. 
The message of Ref.~\cite{vinischmid} is that all diagrams must be calculated, and that normally the tree level diagrams are mostly responsible for the mass distributions, with the effects of the TS diagrams being diluted in these distributions. 

In view of these problems we look now at a related reaction where a TS develops, which is not masked by the effects of tree level and Schmid theorem. 
The reaction is $J/\psi \to \phi \pi^- a_0(980)^+(\pi^+ \eta),\; \phi \pi^+ a_0(980)^-(\pi^- \eta)$. 
The mechanism is similar to that in Figs. 1(a) and 1(b), with $\eta$ replaced by $\phi$, but the intermediate $K\bar{K}$ produce the $a_0(980)$ resonance which decays to $\pi \eta$. 
Then, the tree level diagrams with $K\bar{K}$ production do not interfere with the triangle diagram, which also develops a singularity in the $\pi a_0(980)$ mass distribution. 
It is easy to see where we should expect the singularity by applying Eq. (18) of Ref.~\cite{bayaracetiguo} (with $m_{a_0}$ slightly larger than $2\, m_K$), and one finds that a singularity should appear at $M_{\rm inv}(\pi a_0) \sim 1417 \mev$. 
The purpose of the present work is to do a detailed study of the reaction and make a realistic prediction of the shape and size of the $\pi a_0(980)$ mass distribution in that reaction.

\section{Formalism}
\label{sec:formula}

\begin{figure}[t]
    \centering
    \includegraphics[scale=0.65]{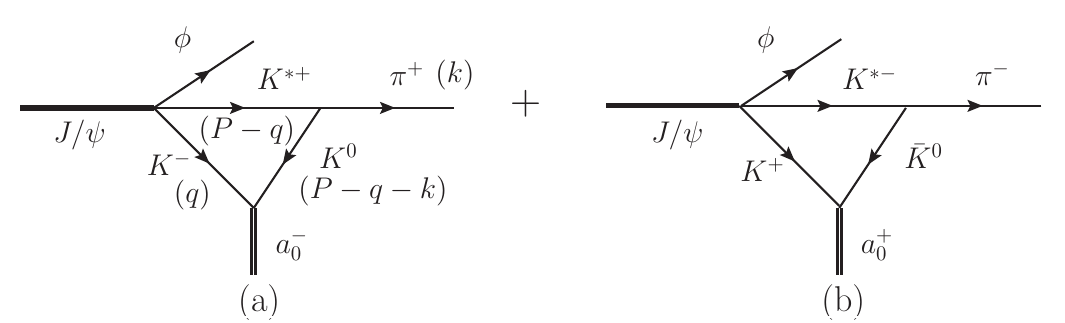}
    \includegraphics[scale=0.68]{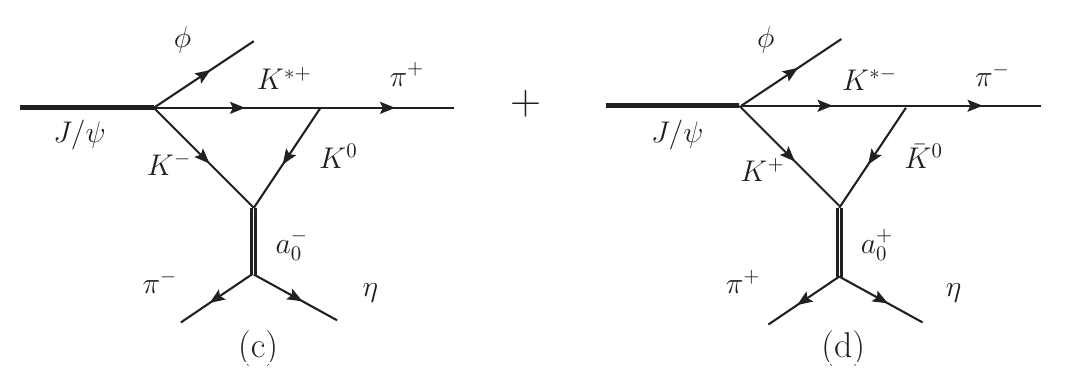}
    \caption{Triangle diagrams for $J/\psi \to \phi \pi^+ a_0^-$ decay (a) and $J/\psi \to \phi \pi^- a_0^+$ decay (b). (c) and (d) illustrate the processes of (a) and (b) respectively, with a clear depiction of the decay channel of $a_0^-$ and $a_0^+$. In (a), the momenta of the particles are shown, where $P=p_{J/\psi} - p_{\phi}$.}
    \label{fig:mech2}
\end{figure}
    
We look at the diagrams of Fig.~\ref{fig:mech2}. 
The diagrams of Figs. \ref{fig:mech2}(a) and \ref{fig:mech2}(b) show the coalescence processes where the $a_0^-,\, a_0^+$ are produced, independent on the decay channel in which the $a_0$ resonances are observed. 
The diagrams of Figs. \ref{fig:mech2}(c) and \ref{fig:mech2}(d) show the explicit reaction, with four body final state when the $a_0^{-/+}$ decay into $\pi^{-/+} \eta$, the only sizeable decay channel. 
We shall consider this final state, but it is practical to consider first the coalescence processes, with only three particles in the final state. 
We consider the reactions $J/\psi \to \phi \pi^+ a_0^-$ and $J/\psi \to \phi \pi^- a_0^+$ as different reactions and will concentrate on the first one. 
The same distributions would be obtained for the second reaction. 
    
In order to be able to determine absolute rates for the $J/\psi \to \phi \pi^+ a_0^- \; (J/\psi \to \phi \pi^- a_0^+)$ reaction we need information on the $J/\psi \to \phi K^{*+} K^-$ reaction, which we take from experiment. 
On the other hand, the dynamics of $K^* \to K \pi$ and $K\bar{K} \to a_0 \to \pi\eta$ are well known, and for the $K\bar{K} \to a_0 \to \pi\eta$ amplitudes we shall use the chiral unitary approach \cite{npa,kaiser,markushin,juan}.

\subsection{The $J/\psi \to \phi K^* \bar{K}$ reaction}
\label{subsec:IIA}

In the Particle Data Group (PDG) \cite{pdg}, we have the branching ratio,
\begin{equation}
\text{Br}(J/\psi \to \phi K^*(892) \bar{K} + c.c.) = (2.18 \pm 0.23)\times 10^{-3}. 
\label{eq:bran}
\end{equation}
However, we are only interested in $J/\psi \to \phi K^{*-} K^+$. 
It is easy to see the rate for this particular channel using isospin and $C$ parity arguments. 
With the isospin convention for the multiplets $(K^+,\, K^0)$, $(\bar{K}^0,\, -K^-)$, $(K^{*+},\, K^{*0})$, $(\bar{K}^{*0},\, -K^{*-})$, the $K^* \bar{K}$ isospin zero states demanded in the reaction of Eq.~\eqref{eq:bran} are given by
\begin{equation}
\begin{aligned}
& | K^* \bar{K},\; I=0 \rangle = -\frac{1}{\sqrt{2}} \;\left( K^{*+} K^- + K^{*0}\bar{K}^0 \right), \\
& | \bar{K}^* K,\; I=0 \rangle = \frac{1}{\sqrt{2}} \;\left( \bar{K}^{*0} K^0 + K^{*-}K^+ \right).
\end{aligned}
\end{equation}
Knowing that $C K^{*+} =- K^{*-}$, $C K^+ = K^-$ etc., the right combination for $J/\psi \to \phi K^* \bar{K} + c.c.$ is given by
\begin{equation}
J/\psi \to \phi \left( K^{*+} K^- + K^{*0}\bar{K}^0 -  K^{*-}K^+ - \bar{K}^{*0} K^0 \right),
\end{equation}
which means that the branching ratio of $J/\psi \to \phi K^{*+} K^-$ will be one fourth of the one in Eq.~\eqref{eq:bran},
\begin{equation}
\text{Br}(J/\psi \to \phi K^{*+} K^-) = (0.55 \pm 0.06)\times 10^{-3}. 
\label{eq:bran2}
\end{equation}
Furthermore, the structure of the amplitude in $S$-wave is given by
\begin{equation}\label{eq:Vam}
t_{J/\psi, \; \phi K^{*+} K^-} = C\; \vec{\epsilon}_{J/\psi} \cdot \left( \vec{\epsilon}_{\phi} \times \vec{\epsilon}_{K^*} \right), 
\end{equation}
with $C$ being a constant. 
We can determine $C$ from the rate of Eq.~\eqref{eq:bran2} using
\begin{equation}
\dfrac{{\rm d} \Gamma_{J/\psi \to \phi K^{*+} K^-}}{{\rm d} M_{\rm inv} (K^{*+} K^-)} = \dfrac{1}{(2\pi)^3} \; \dfrac{1}{4 M_{J/\psi}^2}\; p_{\phi} \, \tilde{p}_{K^-} \,\overline{\sum} \sum |t|^2,
\end{equation}
with 
\begin{equation}
\begin{aligned}
& p_{\phi} = \dfrac{\lambda^{1/2} (M_{J/\psi}^2,\, M_{\phi}^2,\, M_{\rm inv}^2 (K^{*+} K^-))}{2 \,M_{J/\psi}}, \\
& \tilde{p}_{K^-} = \dfrac{\lambda^{1/2} (M_{\rm inv}^2 (K^{*+} K^-),\, M_{K^{*+}}^2, \,M_{K^-}^2)}{2\,M_{\rm inv} (K^{*+} K^-)},
\end{aligned}
\end{equation}
where $\lambda (x,y,z)$ is the K\"{a}ll{\'e}n function defined as $\lambda (x,y,z) = x^2+y^2+z^2-2xy-2xz-2yz$,
and $\overline{\sum} \sum |t|^2$ stands for the average and sum over the polarizations of $J/\psi$, $\phi$, and $K^*$ mesons, with $t$ defined in Eq.~\eqref{eq:Vam}. Therefore,
\begin{equation} \label{eq:2C}
 \overline{\sum} \sum |t|^2 = 2 \,C^2.
\end{equation}
Then we find
\begin{equation}
\dfrac{C^2}{\Gamma_{J/\psi}} = \dfrac{\text{Br}(J/\psi \to \phi K^{*+} K^-)}{\displaystyle\int \dfrac{2}{(2\pi)^3}\; \dfrac{1}{4 M_{J/\psi}^2} \,p_{\phi} \,\tilde{p}_{K^-} \;{\rm d} M_{\rm inv} (K^{*+} K^-)},
\end{equation}
from where we obtain
\begin{equation}\label{eq:FactorC}
\dfrac{C^2}{\Gamma_{J/\psi}} = 1.381 \times 10^{-2} ~~ ( \mev^{-1}),
\end{equation}
which we will use to evaluate the strength of the triangle mechanism.

The structure of the $J/\psi \to \phi K^* \bar K$ vertex is given in Eq.~\eqref{eq:Vam} but it assumes a nonrelativistic reduction.
One can induce such structure from a more general relativistic formulation. 
A suitable operator is
\begin{equation}\label{eq:ref}
  \varepsilon^{\mu\nu\alpha\beta}\, \epsilon_\mu(J/\psi)\,  \epsilon_\nu(\phi)\,\epsilon_\alpha(K^*)\,[a_1\, p_\beta (J/\psi)+a_2\, p_\beta (\phi)+a_3\, p_\beta (K^*)],
\end{equation} 
with $a_i$ independent structures in principle.
Assuming $p/M$ reasonably smaller than $1$ in the average over phase space, the dominant term comes from $\beta=0$, and we have, neglecting the kinetic energies of the $\phi$ and $K^*$,
\begin{equation*}
  \vec \epsilon \,(J/\psi)\cdot [ \vec\epsilon \,(\phi)\times\vec \epsilon \,(K^*)] \,[a_1\, M_{J/\psi}+a_2\, M_{\phi}+a_3\, M_{K^*}],
\end{equation*} 
which means that, independently of the values of $a_i$, there is only one structure in this nonrelativistic limit, the one assumed in Eq.~\eqref{eq:Vam}.
One can estimate the relativistic corrections, assuming just for that purpose that $a_1=a_2=a_3$.
Since $J/\psi$ is at rest, the $\mu$ index in Eq.~\eqref{eq:ref} is spatial.
Hence, the zero index in $\varepsilon^{\mu\nu\alpha\beta}$ can be $\nu$ or $\alpha$ (apart from $\beta$ already considered).
By using
\begin{equation}\label{eq:new}
  \sum_{\rm pol} \epsilon_\mu\; \epsilon_\nu=-g_{\mu\nu}+\frac{p_\mu p_\nu}{M^2},
\end{equation} 
we can look at the interference terms in $\sum_{\rm pol} |t|^2$ between the dominant $\beta =0$ term and the terms with $\nu=0$ or $\alpha =0$, and we find a ratio of the interference term versus the dominant one, ignoring the term of $O(\frac{p^4}{M^4})$,
\begin{equation*}
  -\frac{2}{3} \left( \dfrac{\vec p_{K^*}^{\,2}}{M_{K^*}} -\dfrac{\vec p_{\phi}^{\,2}}{M_{\phi}}\right) \; \dfrac{1}{M_{J/\psi}+M_{K^*}+M_{\phi}},
\end{equation*} 
which basically vanishes over the phase space.
The square of the terms with $\nu=0$ or $\alpha=0$ gives contributions of the order of $O(\frac{p^4}{M^4})$.

On the other hand, Eq.~\eqref{eq:2C} is derived using $\sum_{\rm pol} \epsilon_i \epsilon_j = \delta_{ij}$.
If instead we use Eq.~\eqref{eq:new}, we get some corrections of order $O(\frac{p^2}{M^2})$ relative to the dominant term
\begin{equation}\label{eq:new2}
  \frac{1}{3} \left( \dfrac{\vec p_{K^*}^{\,2}}{M^2_{K^*}} +\dfrac{\vec p_{\phi}^{\,2}}{M^2_{\phi}}\right).
\end{equation} 
We have evaluated the average of Eq.~\eqref{eq:new2} over the phase space for $J/\psi \to \phi K^{*+}K^-$, and we get a correction of $28\%$ over the dominant term.
This uncertainty affects the absolute value of the mass distributions, and it is assumable, since at the end we compare our rates with an experimental one that has $50\%$ error \cite{besthis}.
As a consequence, the formalism is greatly simplified by taking Eqs.~\eqref{eq:Vam} and \eqref{eq:2C}.

One can look at this approximation done from a different perspective. We can assume Eq.~\eqref{eq:Vam}, with the constant $C$ fitted to experiment, as an average of a more complicated structure over the phase space. Then the same average is used in the evaluation of the amplitude of the triangle singularity. Thus this latter decay width, with a similar phase space for $\phi$ and $K^*$,  would be obtained with some relative accuracy.

\subsection{The $a_0^- \to K^-K^0$ coupling and $K^* \to K \pi$ vertex}

The  $K^{*+} \to K^0 \pi^+$ coupling is easily obtained from the standard Lagrangian,
\begin{equation}
{\cal L} = - i g \,\langle [ P,\; \partial_\mu P] V^\mu \rangle,
\end{equation}
with $P$ and $V$ representing the SU(3) $q\bar q$ matrix written in terms of pseudoscalar and vector mesons, respectively \cite{Ikeno:2019grj}.
The coupling $g$ is defined as $g = \frac{M_V}{2f}$, where $M_V=800 \mev$ and  $f=93 \mev$. This yields the vertex
\begin{equation}\label{eq:it}
-i t = -i g \,\epsilon_j (K^*) \;(2 k + q )^j,
\end{equation}
which is evaluated in the frame where we take $\vec{P} = \vec p_{J/\psi} - \vec p_{\phi} = 0$. In this frame, we can neglect the $\epsilon^0$ component of the $K^*$.
Indeed, in that frame, and for $M_{\rm inv}(K^* \bar K) \sim 1417 \mev$ where the TS appears, $p_{K^*} \simeq 150 \mev/c$, and the formula of Appendix A of Ref.~\cite{sakairamos} gives an error by neglecting $\epsilon^0$ of the order of $0.5\%$.

The coupling of $a^-_0 \to K^- K^0$ needed in the evaluation of the diagram of Fig. \ref{fig:mech2}(a) without further decay of $a_0^- \to \pi^- \eta$ can be accounted for in the following way. 
It is clear that if we evaluate the $J/\psi \to \phi \pi^+ \pi^- \eta$ decay we would only need the $K^- K^0 \to \pi^- \eta$ amplitude. 
The coalescence decay $J/\psi \to \phi \pi^+ a_0^-$ should also be able to be calculated using the $K\bar{K}$ amplitudes and this is formally done as discussed below.

Assuming that, close to the peak of the $a_0(980)$, 
\begin{equation}
t_{K^- K^0,\, K^-K^0} (M_{\rm inv}) = \frac{g^2_{a_0,\, K^-K^0}}{M^2_{\rm inv} - m^2_{a_0} + i M_{\rm inv} \Gamma_{a_0}},
\end{equation}
with $\Gamma_{a_0}$ considered constant for the formal derivation,
then, using Cauchy's integration we find immediately
\begin{equation}
g^2_{a_0,\, K^-K^0} = -\frac{1}{\pi} \int {\rm d} M^2_{\rm inv} \; \text{Im}\, t_{K^- K^0,\, K^-K^0} (M_{\rm inv}), 
\label{eq:ga0}
\end{equation}
which is also trivially obtained in the limit of $\Gamma_{a_0} \to 0$ using
\begin{equation}
\text{Im}\, \left[ \left( M^2_{\rm inv} - m^2_{a_0} \right) + i \varepsilon \right]^{-1} = - \pi \delta \left( M^2_{\rm inv} - m^2_{a_0} \right).
\end{equation}
In the coalescence process we will use Eq.~\eqref{eq:ga0} in the evaluation of the $|t_{\rm TS}|^2$ of the triangle diagram. 
We will have
\begin{equation}
\dfrac{{\rm d } \Gamma_{J/\psi \to \phi \pi^+ a_0^-}}{{\rm d} M_{\rm inv} (\pi^+ a_0^-)} = \dfrac{1}{(2\pi)^3} \; \dfrac{1}{4 \,M_{J/\psi}^2} \;p_{\phi} \;\tilde{p}_{\pi^+} \;\overline{\sum} |t_{\rm TS}|^2,
\label{eq:dgamJ}
\end{equation}
where
\begin{align}
& p_{\phi} = \dfrac{\lambda^{1/2} (M_{J/\psi}^2, \,m_{\phi}^2, \,M_{\rm inv}^2 (\pi^+ a_0^-))}{2 \,M_{J/\psi}}, \\[2mm]
& \tilde{p}_{\pi^+} = \dfrac{\lambda^{1/2} (M_{\rm inv}^2 (\pi^+ a_0^-),\, m_{\pi^+}^2, \,m_{a_0^-}^2)}{2\,M_{\rm inv} (\pi^+ a_0^-)}, \label{eq:tilpi}
\end{align}
with $t_{\rm TS}$ corresponding to the amplitude for the process of Fig.~\ref{fig:mech2}(a).
Since $|t_{\rm TS}|^2$ contains $g^2_{a_0,\, K^-K^0}$, given in Eq.~\eqref{eq:ga0}, we can undo the ${\rm d} M^2_{\rm inv}$ integration in Eq.~\eqref{eq:dgamJ} and write
\begin{equation}
\begin{aligned}
\dfrac{{\rm d}^2 \Gamma_{J/\psi\to  \phi \pi^+ a_0^- (\pi^- \eta)}}{{\rm d} M_{\rm inv} (\pi^- \eta) \;{\rm d} M_{\rm inv} (\pi^+ a_0^-)} =& -\dfrac{1}{\pi} \;2 M_{\rm inv}(\pi^- \eta) \;\text{Im}\, t_{K^- K^0,\, K^-K^0} (M_{\rm inv} (\pi^- \eta)) \\[2mm]
&\times \dfrac{1}{(2\pi)^3}\; \dfrac{1}{4 \,M_{J/\psi}^2} \; p_{\phi} \; \tilde{p}_{\pi^+}^{\,\prime} \overline{\sum} |\tilde{t}_{\rm TS}|^2,
\end{aligned}
\label{eq:dgamJ2}
\end{equation}
where $\tilde{p}_{\pi^+}^{\,\prime}$ is given by Eq.~\eqref{eq:tilpi} substituting $m_{a_0^-}$ by $M_{\rm inv} (\pi^- \eta)$, and $|\tilde{t}_{\rm TS}|^2$ is the magnitude $|t_{\rm TS}|^2$ where we remove $g^2_{a_0,\, K^-K^0}$.

Eq.~\eqref{eq:dgamJ2} can be immediately reinterpreted. 
Indeed, in the evaluation of $|t_{\rm TS}|^2$ for the triangle diagram of Fig. \ref{fig:mech2}(c), we will need $|t_{K^- K^0,\; \pi^-\eta}|^2$ along with  the extra phase space for $\pi^-\eta$ with respect to Fig.~\ref{fig:mech2}(a). 
But $|t_{K^- K^0,\; \pi^-\eta}|^2$ times the phase space for $a_0^- \to \pi^- \eta$ decay is what is given by $\text{Im}\, t_{K^- K^0,\; K^-K^0}$ via the optical theorem. 
The derivation done, however, has  served to go from the differential mass distribution in the three body final state to the one of the four body in a simple and intuitive way. 
There is a caveat, however, since above the $K\bar K$ threshold $\text{Im}\, t_{K^- K^0,\; K^-K^0}$, via the optical theorem, also contains the decay of $a_0$ into $K\bar K$.
However, close to the $K\bar K$ threshold where we move, this decay rate is very small and we can safely rely on Eq.~\eqref{eq:dgamJ2} for our purposes. \footnote{The PDG \cite{pdg} reports $\Gamma(K\bar K)/\Gamma(\pi \eta) \simeq 0.17$, but a relative large span of energies above the $K\bar K$ threshold is considered to get that number.}

\subsection{The triangle amplitude}

In Fig. \ref{fig:mech2}(a), we show the momenta of the particles. 
Since we are concerned about the triangle singularity, occurring when the intermediate particles are all on shell, we can simplify the calculation (see Ref.~\cite{bayaracetiguo}) by taking the positive energy part of the propagators, which is  the one that can go on shell. 
Hence for the $K^-$ propagator we would write
\begin{equation}
\dfrac{1}{q^2 - m_K^2 + i\varepsilon} = \dfrac{1}{2\,\omega(\vec q \,)} \;\left( \dfrac{1}{q^0 - \omega_K(\vec q\,) + i\varepsilon} - \dfrac{1}{q^0 + \omega_K(\vec q\,) -i\varepsilon} \right),
\end{equation}
with $\omega_K(\vec q\,) = \sqrt{\vec{q}\;^2 + m_K^2}$, and with $q^0$ positive only the first term of the former equation can go on shell, and we shall then keep this term alone. 
This simplifies the expression for the loop amplitude which reads, removing the $g_{a_0,\, K^-K^0}$ vertex,
\begin{equation}
\begin{aligned}
-i \tilde{t}_{\rm TS} =& -i\, C \int \dfrac{{\rm d}^4 q}{(2\pi)^4} \;\varepsilon_{ijl}\; \epsilon_i(J/\psi) \;\epsilon_j(\phi) \;\epsilon_l(K^*)\; (-i) g \;\epsilon_m(K^*) \;(2k + q)_m  \\[2.5mm]
&\times (-i)\;\dfrac{1}{2\,\omega_{K^-}(\vec q\,)}\; \dfrac{1}{2\,\omega_{K^0}(\vec{q} + \vec{k})}\; \dfrac{1}{2\,\omega_{K^{*+}}(\vec q\,)}\; \dfrac{i}{q^0 - \omega_{K^-}(\vec q\,) + i\varepsilon}  \\[2.5mm]
&\times \frac{i}{P^0 - q^0 - \omega_{K^{*+}}(\vec q \,) + i \frac{\Gamma_{K^*}}{2}} \; \dfrac{i}{P^0 - q^0 - k^0 - \omega_{K^0}(\vec{q} + \vec{k}\,) + i\varepsilon},
\end{aligned}
\label{eq:loop1}
\end{equation}
with $\omega_{K^-}(\vec q \,) = \sqrt{\vec{q}\;^2 + m_{K^-}^2}$, and $\omega_{K^{*+}}(\vec q\,) = \sqrt{\vec{q}\;^2 + m_{K^{*}+}^2}$, where we have explicitly taken into account the $K^*$ width in the $K^*$ propagator, and $P^0$, $k^0$ are given by
\begin{equation}
\begin{aligned}
& P^0 = M_{\rm inv} (\pi^+ a_0^-), \\[2mm]
& k^0 = \dfrac{P^{02} + m^2_{\pi^+} - M^2_{\rm inv} (\pi^- \eta)}{2 P^0}.
\end{aligned}
\end{equation}
Since we know that the TS gets its strength from placing the intermediate particles of the loop on shell, we can rely upon the arguments used after Eq.~\eqref{eq:FactorC} and after Eq.~\eqref{eq:it} to keep only the spatial components of the polarization vectors of the vector mesons.
We sum over the $K^*$ polarizations in the loop, $\sum_{\rm pol} \epsilon_l(K^*) \epsilon_m(K^*) = \delta_{lm}$, 
good for the $K^*$ with small momenta that we have in the TS region,
and integrating analytically over $q^0$ in Eq.~\eqref{eq:loop1} using Cauchy's residues, we obtain
\begin{equation}
\begin{aligned}
\tilde{t}_{\rm TS} =& g\, C\, \varepsilon_{ijl}\; \epsilon_i(J/\psi) \; \epsilon_j(\phi) \; \int \dfrac{{\rm d}^3 q}{(2\pi)^3} \;(2k + q)_l\; \dfrac{1}{2\,\omega_{K^-}(\vec q\,)}\; \dfrac{1}{2\,\omega_{K^{*+}}(\vec q\,)}\; \dfrac{1}{2\,\omega_{K^0}(\vec{q} + \vec{k}\,)}  \\[2.5mm]
&\times \dfrac{i}{P^0 - \omega_{K^-}(\vec q\,) - \omega_{K^{*+}}(\vec q\,) + i \frac{\Gamma_{K^*}}{2}}\; \dfrac{i}{P^0 - k^0- \omega_{K^-}(\vec q\,) - \omega_{K^0}(\vec{q} + \vec{k}\,) + i\varepsilon}.
\end{aligned}
\end{equation}

Next, considering that
\begin{equation}
\int {\rm d}^3 q\, F(\vec{q}, \vec{k})\, q_l = k_l\, \int {\rm d}^3 q\; F(\vec{q}, \vec{k})\; \dfrac{\vec{q} \cdot \vec{k}}{\vec{k}^{\,2}},
\end{equation}
we finally write $\tilde{t}_{\rm TS}$ as
\begin{equation}
\tilde{t}_{\rm TS} = g\, C\, \epsilon_{ijl}\; \epsilon_i(J/\psi) \; \epsilon_j(\phi)\; k_l \; \tilde{t}'_{\rm TS},
\end{equation}
with
\begin{equation}
\begin{aligned}
  \tilde{t}'_{\rm TS} =& \int \dfrac{{\rm d}^3 q}{(2\pi)^3}\; \theta(q_{max} - |\vec{q}\;^*|\,)\; \left(2 + \dfrac{\vec{q} \cdot \vec{k}}{\vec{k}^{\,2}}\right)\; \dfrac{1}{2\,\omega_{K^-}(\vec q\,)}\, \dfrac{1}{2\,\omega_{K^{*+}}(\vec q\,)}\, \dfrac{1}{2\,\omega_{K^0}(\vec{q} + \vec{k}\,)}  \\[2.5mm]
&\times \dfrac{i}{P^0 - \omega_{K^-}(\vec q\,) - \omega_{K^{*+}}(\vec q\,) + i \dfrac{\Gamma_{K^*}}{2}}\; \dfrac{i}{P^0 - k^0- \omega_{K^-}(\vec q\,) - \omega_{K^0}(\vec{q} + \vec{k}\,) + i\varepsilon},
\end{aligned}
\label{eq:loop2}
\end{equation}
and 
\begin{equation}
\overline{\sum} \left| \tilde{t}_{\rm TS} \right|^2 = \frac{2}{3}\, \vec{k}^{\,2}\, g^2\, C^2\; \left| \tilde{t}'_{\rm TS} \right|^2 .
\end{equation}
In Eq.~\eqref{eq:loop2} we have introduced the factor $\theta(q_{max} - |\vec{q}\;^*|)$, where $\vec{q}\;^*$ is the $K^-$ momentum in the $\pi^- \eta$ rest frame given by \cite{bayaracetiguo}
\begin{equation}
\vec{q}\,^* = \left[ \left( \frac{E_{a_0}}{M_{\rm inv} (\pi^- \eta)} - 1 \right) \frac{\vec{q} \cdot \vec{k}}{\vec{k}^{\,2}} + \frac{q^0}{M_{\rm inv} (\pi^- \eta)} \right] \vec{k} + \vec{q}\; ,
\end{equation}
with $E_{a_0} = \sqrt{M^2_{\rm inv} + \vec{k}^{\,2}}$, and $q^0 = \sqrt{m^2_K + \vec{q}^{\;2}}$. 
The $\theta(q_{\rm max}-|\vec q^{\,*}|)$ factor is needed in Eq.~\eqref{eq:loop2} and is justified as follows.
The chiral unitary approach relies upon the solution of the Bethe-Salpeter equation with coupled channels.
The usual on-shell factorization of the loops can be justified using dispersion relations as in Ref.~\cite{OllerUlf}, 
but also in a simpler way that we outline below.
Following Ref.~\cite{danijuan} we start from a separable potential in momentum space of the type
\begin{equation*}
  V(\vec q, \vec q\,') = V \theta(q_{\rm max}-|\vec q\,|)\, \theta(q_{\rm max}-|\vec q\,'|),
\end{equation*} 
from where we can see the meaning of $q_{\rm max}$ as measuring the range of the interaction in momentum space.
If one constructs the $T$ matrix from this potential,
\begin{equation*}
 T(\vec q, \vec q\,') = V(\vec q, \vec q\,') + i \int \dfrac{{\rm d}^4 p}{(2\pi)^4}\; V(\vec q, \vec p\,) \; \dfrac{1}{p^2-m_1^2+i\varepsilon} \; \dfrac{1}{(P-p)^2 -m_2^2+i \varepsilon} \; T(\vec p, \vec q\,'),
\end{equation*}
with $P$ the total momentum and $m_1, m_2$ the masses of the intermediate particles, one can immediately see that $T(\vec q, \vec q\,')$ is also separable as
\begin{equation*}
 T(\vec q, \vec q\,') = T \; \theta(q_{\rm max}-|\vec q\,|)\, \theta(q_{\rm max}-|\vec q\,'|),
\end{equation*}
and
\begin{equation*}
  T= \dfrac{V}{1-VG},
\end{equation*}
with $G$, after performing analytically the $p^0$ integration, given by
\begin{equation*}
  G= \int_{|\vec{p}\,| < q_{\rm max}} \dfrac{{\rm d}^3 p}{(2\pi)^3}\;\dfrac{\omega_1(\vec p\,) + \omega_2(\vec p\,)}{2\,\omega_1(\vec p\,)\, \omega_2(\vec p\,)} \; \dfrac{1}{{P^0}^2-[\omega_1(\vec p\,) + \omega_2(\vec p\,)]^2 + i\varepsilon},
\end{equation*}
with $\omega_i(\vec p\,)= \sqrt{\vec p^{\;2}+m_i^2}$.
The value of $q_{\rm max}$ is taken from Ref.~ \cite{Liang:2014tia,lianglin} as $q_{\rm max} = 600 \mev/c$.

\subsection{$K\bar{K}$ amplitudes}

We need to calculate the $t_{K^-K^0,\; K^-K^0}$ amplitude, for which we use the chiral unitary approach.
In this case, the $t_{K^- K^0 , K^- K^0}$ amplitude is one of the $T$-matrix elements, that is extracted by solving the Bethe-Salpeter equation in coupled-channels,  
\begin{equation}\label{eq:BS}
T = [ 1 - V G]^{-1} V ,
\end{equation}
with $V$ the kernel encoding the $V_{ij}$ amplitudes from the $i$ channel to the $j$ channel. The channels considered are $K\bar{K}$, $\pi \eta$, $\pi \pi$, and $\eta \eta$.
The relevant $V_{ij}$ amplitudes among all the channels in our case are taken from Ref. \cite{lianglin}, using explicitly the $\eta-\eta'$ mixing of Ref.~\cite{bramon} [Eq.~(A.4) of Ref. \cite{lianglin}]. 
Furthermore, $G$ is a diagonal matrix with its elements $G_l$ corresponding to the loop function for the $l$th channel.
We take the $G_l$ loop function regularized by means of a cutoff $q_{\rm max}$ in the three-momentum,
\begin{equation}
G_l = \int_{|\vec{q}\,| < q_{\rm max}} \dfrac{{\rm d}^3 q}{(2\pi)^3}\; \dfrac{\omega_1 + \omega_2}{2\,\omega_1\, \omega_2}\; \dfrac{1}{s-(\omega_1 + \omega_2)^2 + i\varepsilon}.
\end{equation}

Since, only the zero charged components are considered in Ref. \cite{lianglin}, we make use of the fact that $K^-K^0$ is the $I=1,\, I_3=-1$ component of $K\bar{K}$ and write
\begin{equation}
t_{K^-K^0,\; K^-K^0} = \frac{1}{2} \left( t_{K^0 \bar{K}^0,\; K^0 \bar{K}^0} + t_{K^+K^-,\; K^+K^-} - 2\, t_{K^0 \bar{K}^0,\; K^+K^-} \right).
\end{equation}
Following Refs.~\cite{Liang:2014tia,lianglin}, we take $q_{\rm max} = 600 \mev/c$.

\section{Results} 

\begin{figure}[t]
    \begin{center}
    \includegraphics[scale=0.45]{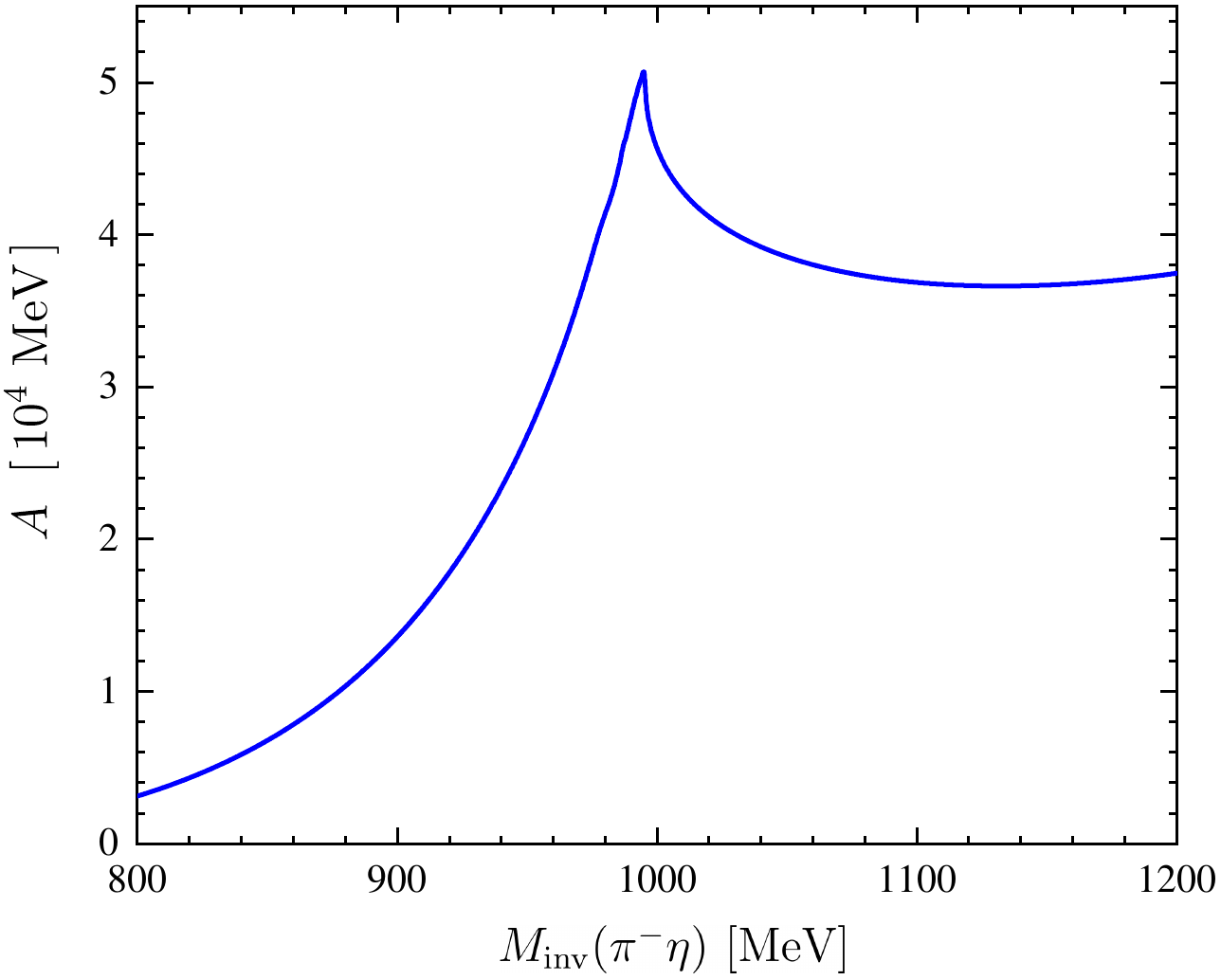}
    \end{center}
    \vspace{-0.7cm}
    \caption{Factor $A \equiv\big[- \dfrac{2}{\pi}\; M_{\rm inv}(\pi^- \eta) \; {\rm Im}\, t_{K^- K^0, K^-K^0} (M_{\rm inv}(\pi^- \eta))\big]$ as a function of $M_{\rm inv}(\pi^- \eta)$.}
    \label{Fig:tKK2}
\end{figure}

In Fig.~\ref{Fig:tKK2}, we show the factor $A \equiv [- \dfrac{2}{\pi}\; M_{\rm inv}(\pi^- \eta) \; {\rm Im}\, t_{K^- K^0, K^-K^0} (M_{\rm inv}(\pi^- \eta))]$, which is involved in the double differential decay width of Eq.~\eqref{eq:dgamJ2}.
We can see a cusp-like structure around $M_{\rm inv}(\pi^- \eta) =m_{a_0} = 980\, \mev$, reflecting the spectral function of the $a_0(980)$.
The shape of Fig.~\ref{Fig:tKK2} is interesting and it does not reflect $|t_{K\bar K, \pi\eta}|^2$, because, through the optical theorem, ${\rm Im}\, t_{K^- K^0, K^-K^0}$ contains a part from the transition of $K^-K^0$ to $\pi^- \eta$ in which we are interested, and also $K^- K^0 \to K\bar K$.
This is the reason for the flattening of the $A$ factor as we go away from the $K\bar K$ threshold.
One guarantees to keep only the $K\bar K \to \pi \eta$ transition in a region up to $\sim 1050\mev$.

Next, we discuss the amplitude $\tilde{t}'_{\rm TS}$ of Eq.~\eqref{eq:loop2} and $\frac{1}{\Gamma_{J/\psi}}\,\frac{{\rm d}^2 \Gamma_{J/\psi \to \phi \pi^+ a_0(980)^-}}{{\rm d} M_{\rm inv}(\pi^- \eta) \;{\rm d} M_{\rm inv}(\pi^+ a_0^-)}$ of Eq.~\eqref{eq:dgamJ2}, 
which are functions of both $M_{\rm inv}(\pi^- \eta)$ and $M_{\rm inv}(\pi^+ a_0^-)$. 
We will present the results in three cases: 1) fixing $M_{\rm inv}(\pi^- \eta)=m_{a_0}=980\mev$; 2) fixing $M_{\rm inv}(\pi^+ a_0^-)=1416 \,\mev$, where the TS occurs in the triangle loops of Fig.~\ref{fig:mech2}; and 3) integrating $\frac{1}{\Gamma_{J/\psi}}\,\frac{{\rm d}^2 \Gamma_{J/\psi \to \phi \pi^+ a_0(980)^-}}{{\rm d} M_{\rm inv}(\pi^- \eta) \;{\rm d} M_{\rm inv}(\pi^+ a_0^-)}$ over $M_{\rm inv}(\pi^- \eta)$.  

\subsection{Fixing $M_{\rm inv}(\pi^- \eta)=m_{a_0}=980\mev$}
\vspace{-0.5cm}

\begin{figure}[t]
  \begin{center}
  \includegraphics[scale=0.48]{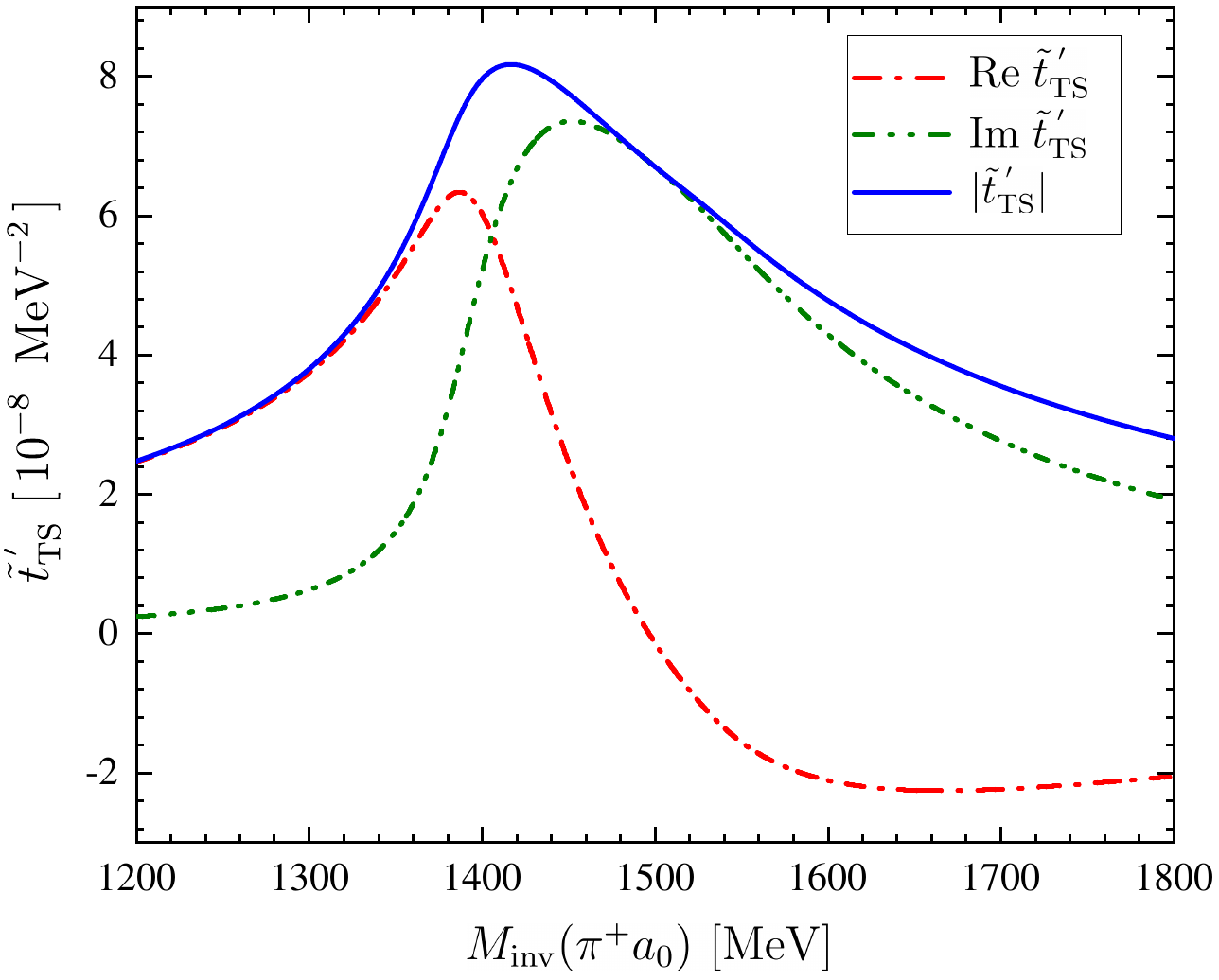}
  \end{center}
  \vspace{-0.7cm}
  \caption{$\tilde{t}'_{\rm TS}$ given by Eq.~\eqref{eq:loop2} as a function of $M_{\rm inv}(\pi^+ a_0^-)$ when fixing $M_{\rm inv}(\pi^- \eta)=m_{a_0}$.}
  \label{Fig:TS1}
  \end{figure}

In Fig.~\ref{Fig:TS1}, we show results for ${\rm Re}\,\tilde{t}'_{\rm TS}$,  ${\rm Im}\,\tilde{t}'_{\rm TS}$ and $|\tilde{t}'_{\rm TS}|$ as a function of $M_{\rm inv}(\pi^+ a_0^-)$ while keeping $M_{\rm inv}(\pi^- \eta)$ fixed.
The structure of the amplitude exhibits features typical of triangle singularities observed in other cases (see Fig.~5 of Ref.~\cite{sakairamos}, Fig.~4 of Ref.~\cite{PavaoSakai}, Fig.~3 of Ref.~\cite{sakailiang}, Fig.~4 of Ref.~\cite{OsetRoca}, Fig.~5 of Ref.~\cite{DaiYu}, 
Fig.~4 of Ref.~\cite{LiangChen}, and Fig.~19 of Ref.~\cite{GuoRev}).
It has the imaginary part peaking around the $M_{\rm inv}(\pi^+ a_0^-)$, as provided by Eq.~(18) of Ref.~\cite{bayaracetiguo}, and the real part changes sign around the peak of the imaginary part.
It resembles much the structure of a resonance, and hence the danger to identify these peaks as genuine resonances, but as we can see, the origin of this structure simply comes from the triangle diagram and is exclusively tied to the combination of masses and invariant masses of the particles involved.
It does not have its origin in the interaction of quarks or the interaction of hadrons. This is why it is referred to as a kinematical singularity.
The modulus of this amplitude, $|\tilde{t}'_{\rm TS}|$, has a clear peak that should manifest in the studied reaction.

\begin{figure}[t]
  \begin{center}
  \includegraphics[scale=0.48]{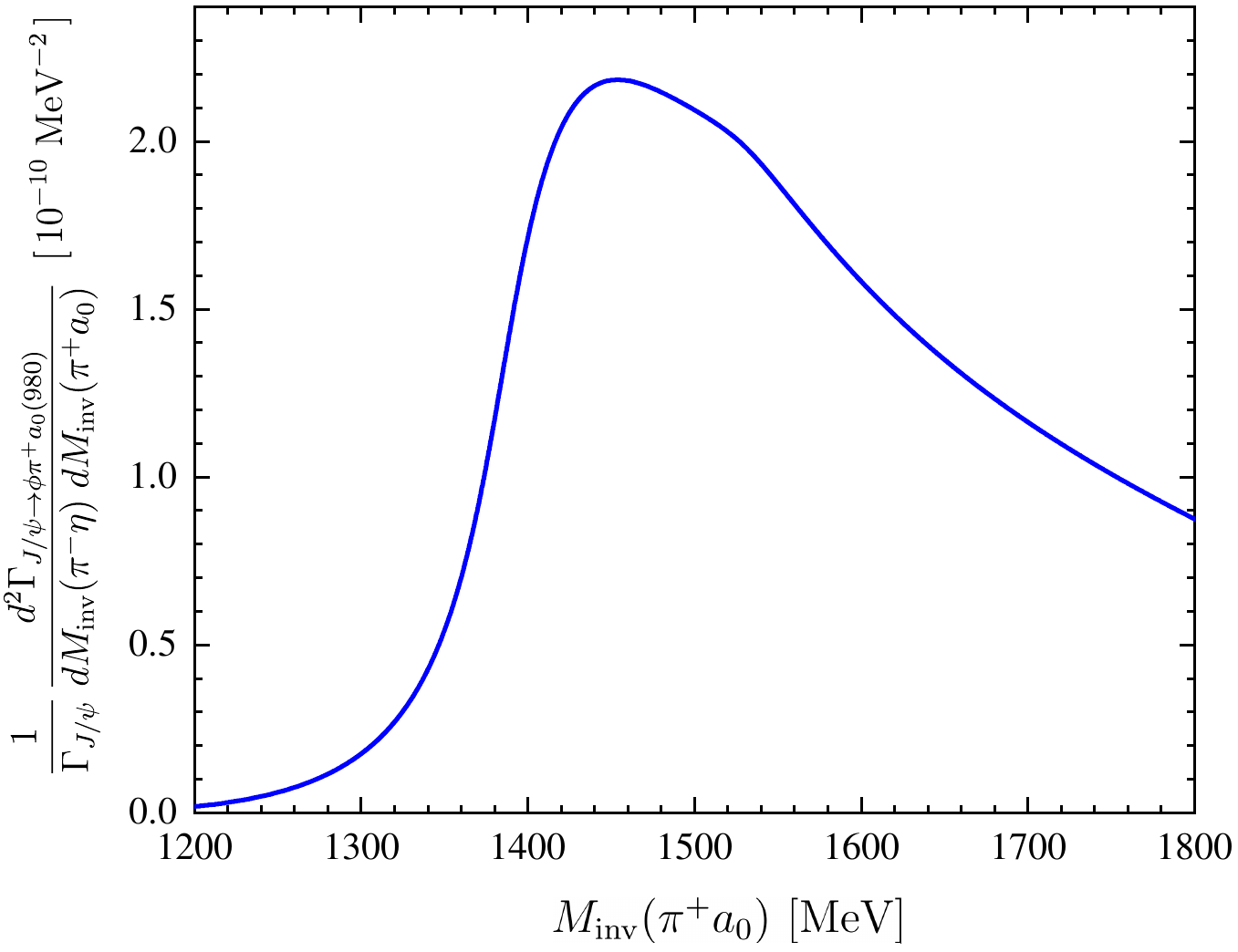}
  \end{center}
  \vspace{-0.7cm}
  \caption{$\frac{1}{\Gamma_{J/\psi}}\,\frac{{\rm d}^2 \Gamma_{J/\psi \to \phi \pi^+ a_0(980)^-}}{{\rm d} M_{\rm inv}(\pi^- \eta) \;{\rm d} M_{\rm inv}(\pi^+ a_0^-)}$ as a function of $M_{\rm inv}(\pi^+ a_0^-)$ when fixing $M_{\rm inv}(\pi^- \eta)=m_{a_0}$.}
  \label{Fig:dMinv1}
\end{figure}

In Fig.~\ref{Fig:dMinv1}, we show the double differential decay width normalized to the $J/\psi$ width as a function of $M_{\rm inv}(\pi^+ a_0^-)$, while fixing $M_{\rm inv}(\pi^- \eta)=m_{a_0}=980\mev$.
We see a clear peak around $M_{\rm inv}(\pi^+ a_0^-) =1440 \mev$, coming from $|\tilde{t}'_{\rm TS}|$ in Eq.~\eqref{eq:dgamJ2}.
This is what one would observe in a devoted experiment with a bin of $1 \mev$ for $M_{\rm inv}(\pi^+ a_0^-)$, and $1 \mev$ for $M_{\rm inv}(\pi^- \eta)$.
Obviously, the accumulation of events in bigger bins increases the statistics, as we shall see.

\subsection{Fixing $M_{\rm inv}(\pi^+ a_0^-)=1416 \,\mev$}

\begin{figure}[t]
  \begin{center}
  \includegraphics[scale=0.48]{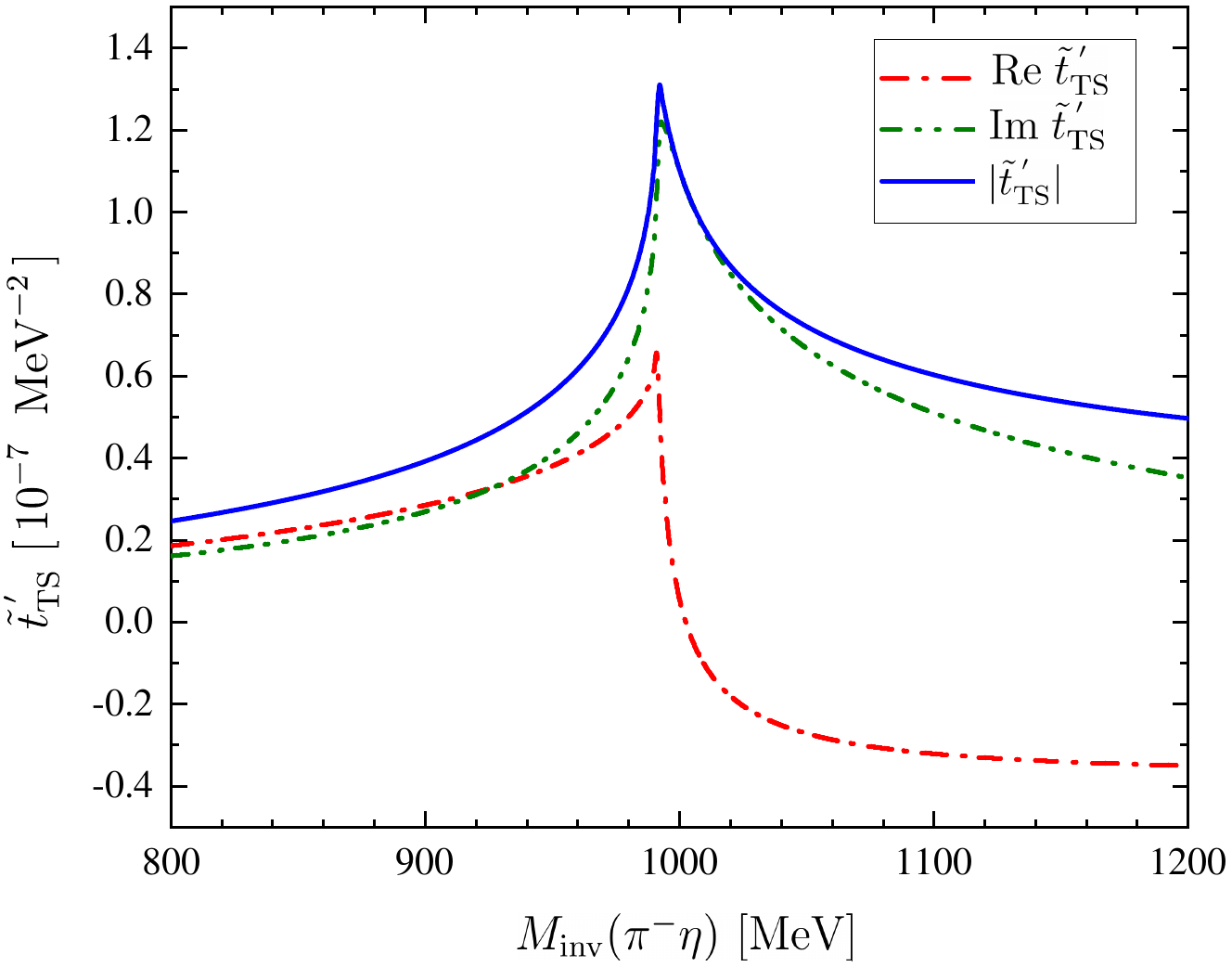} 
  \end{center}
  \vspace{-0.7cm}
  \caption{$\tilde{t}'_{\rm TS}$ given by Eq.~\eqref{eq:loop2} as a function of $M_{\rm inv}(\pi^- \eta)$ when fixing $M_{\rm inv}(\pi^+ a_0^-)=1416 \mev$.}
  \label{Fig:tTSmid}
\end{figure}

In Fig.~\ref{Fig:tTSmid}, we now set $M_{\rm inv}(\pi^+ a_0^-)=1416 \,\mev$ and plot $\tilde{t}'_{\rm TS}$ as a function of $M_{\rm inv}(\pi^- \eta)$. 
Once more, we display the real and imaginary parts of the amplitude, as well as its modulus.
We see again that the imaginary part and the modulus delineate the shape of the $a_0(980)$ resonance.
The real part changes sign at the peak of the $a_0(980)$, reflecting a typical resonance behavior.
It is interesting to see that even if the $a_0(980)$ appears as a cusp, corresponding to a nearly missed bound state, or virtual state, it still exhibits the typical shape of a resonance amplitude. 
Such kinds of behaviors for nearly missed bound states can be seen in other cases.
For instance, in the $p d \to {^3\rm He} \,\eta$ reaction \cite{He3eta} (see Fig.~8 of that work), the amplitude exhibits a resonance structure with a peak very close to and just below the ${^3\rm He} \,\eta$ threshold. However, there is no pole below threshold, and technically, no bound state.

\begin{figure}[t]
\begin{center}
\includegraphics[scale=0.45]{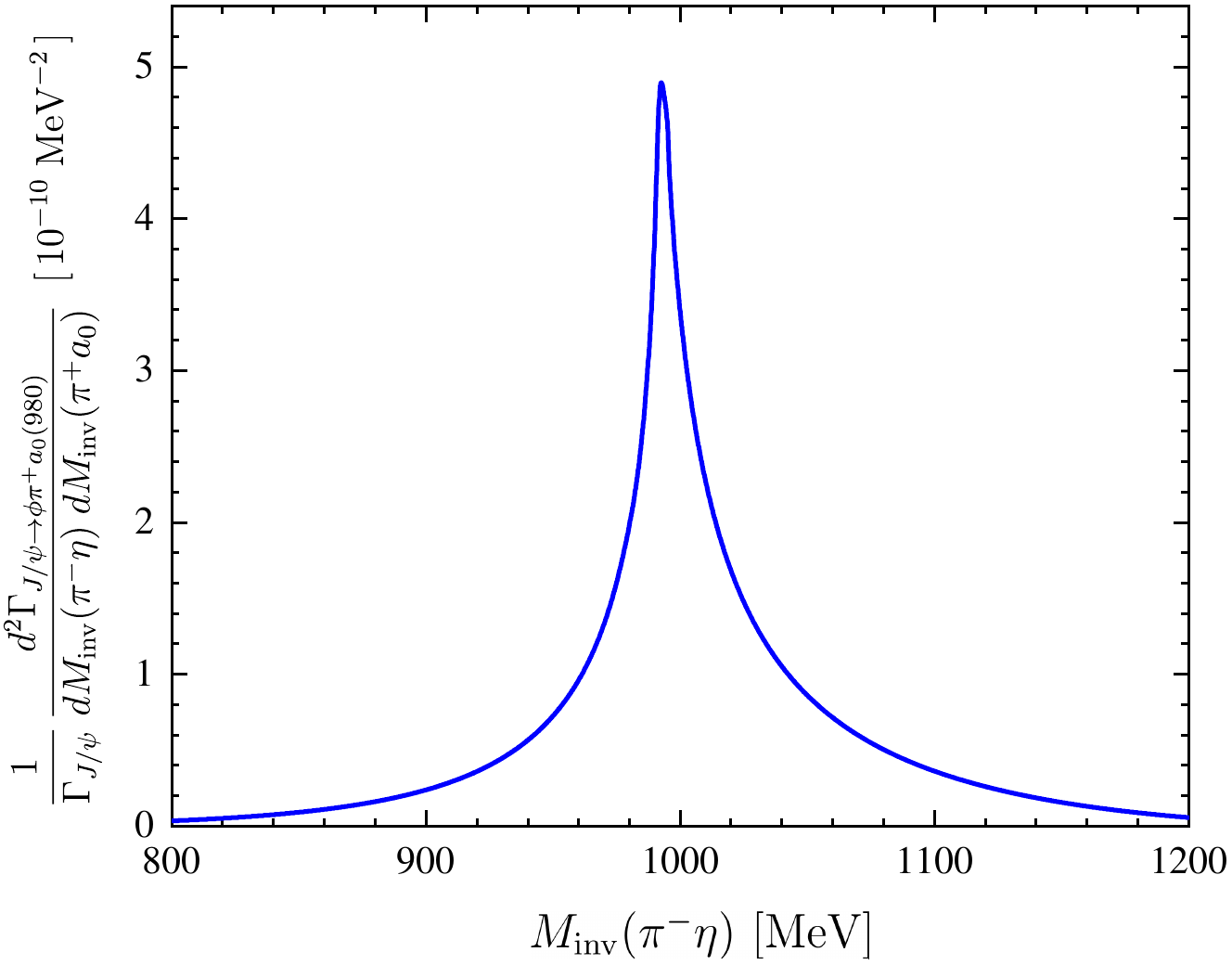}
\end{center}
\vspace{-0.7cm}
\caption{$\frac{1}{\Gamma_{J/\psi}}\,\frac{{\rm d}^2 \Gamma_{J/\psi \to \phi \pi^+ a_0(980)^-}}{{\rm d} M_{\rm inv}(\pi^- \eta) \;{\rm d} M_{\rm inv}(\pi^+ a_0^-)}$ as a function of $M_{\rm inv}(\pi^- \eta)$ when fixing $M_{\rm inv}(\pi^+ a_0^-)=1416\mev$.}
\label{Fig:Gama0Mid}
\end{figure}

In Fig.~\ref{Fig:Gama0Mid}, we show again $\frac{1}{\Gamma_{J/\psi}}\,\frac{{\rm d}^2 \Gamma_{J/\psi \to \phi \pi^+ a_0(980)^-}}{{\rm d} M_{\rm inv}(\pi^- \eta) \;{\rm d} M_{\rm inv}(\pi^+ a_0^-)}$ as a function of $M_{\rm inv}(\pi^- \eta)$, fixing now  $M_{\rm inv}(\pi^+ a_0^-)$ at the peak of the TS amplitude.
This comes from Eq.~\eqref{eq:dgamJ2} and contains $|\tilde{t}'_{\rm TS}|^2$ along with the phase space.
The shape of the $a_0(980)$ resonance shows up as a clear cusp structure, as seen in recent experiments with high resolution \cite{Rubin,midhalo,besnew}.
  
\subsection{Integrating $\frac{1}{\Gamma_{J/\psi}}\,\frac{{\rm d}^2 \Gamma_{J/\psi \to \phi \pi^+ a_0(980)^-}}{{\rm d} M_{\rm inv}(\pi^- \eta) \;{\rm d} M_{\rm inv}(\pi^+ a_0^-)}$ over $M_{\rm inv}(\pi^- \eta)$}

\begin{figure}[t]
    \begin{center}
    \includegraphics[scale=0.5]{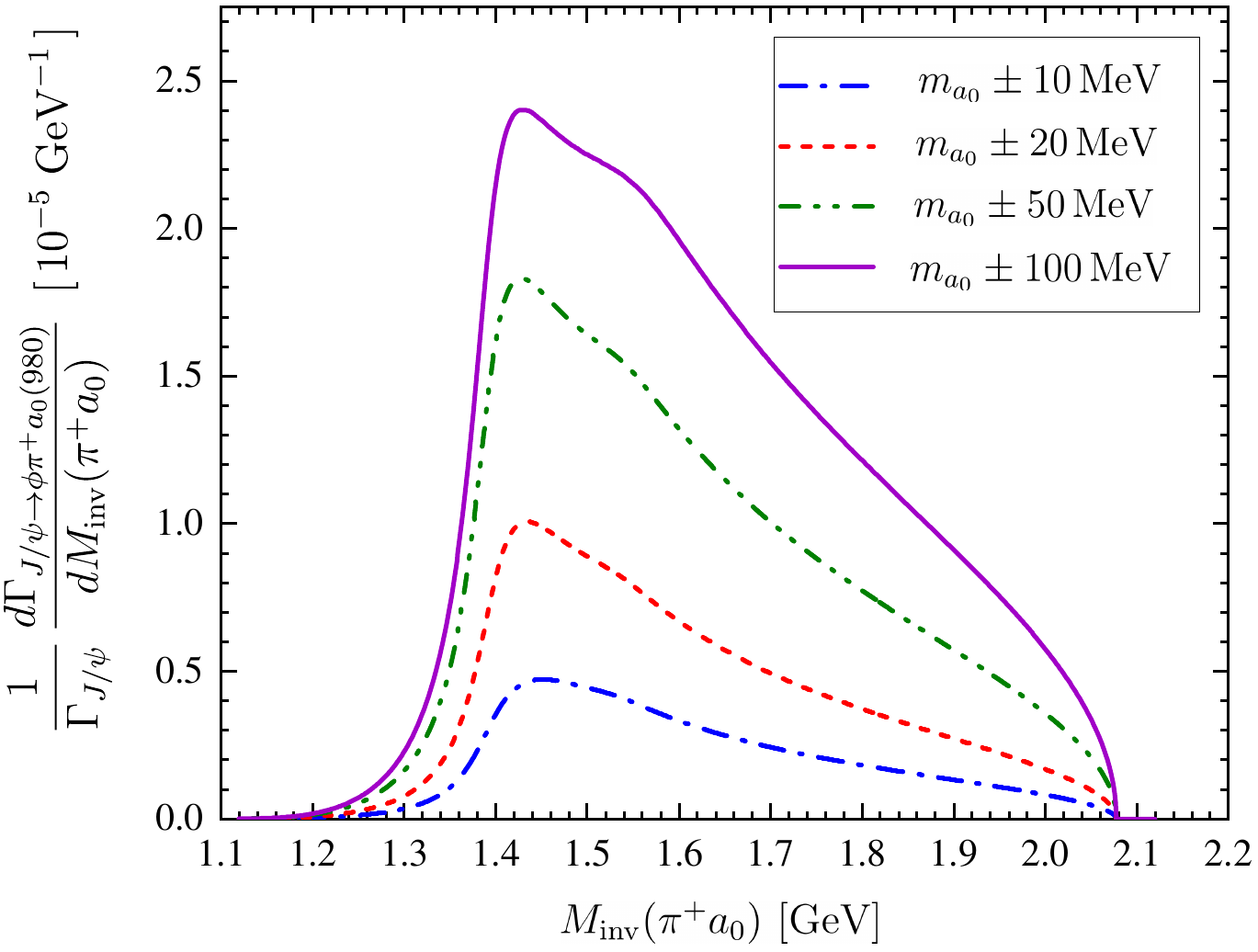}
    \end{center}
    \vspace{-0.7cm}
    \caption{$\frac{1}{\Gamma_{J/\psi}}\,\frac{d^2 \Gamma_{J/\psi \to \phi \pi^+ a_0(980)^-}}{{\rm d} M_{\rm inv}(\pi^- \eta) \;{\rm d} M_{\rm inv}(\pi^+ a_0^-)}$ as a function of $M_{\rm inv}(\pi^+ a_0^-)$ when integrating over $M_{\rm inv}(\pi^- \eta)$ in the ranges: $m_{a_0} \pm 10 \mev$, $m_{a_0} \pm 20 \mev$, $m_{a_0} \pm 50 \mev$ and $m_{a_0} \pm 100 \mev$.}
    \label{Fig:Minv3}
 \end{figure}

Fig.~\ref{Fig:Minv3} shows $\frac{1}{\Gamma_{J/\psi}}\,\frac{{\rm d}^2 \Gamma_{J/\psi \to \phi \pi^+ a_0(980)^-}}{{\rm d} M_{\rm inv}(\pi^- \eta) \;{\rm d} M_{\rm inv}(\pi^+ a_0^-)}$ as a function of $M_{\rm inv}(\pi^+ a_0^-)$, when integrating over $M_{\rm inv}(\pi^- \eta)$ in the ranges of $m_{a_0} \pm 10 \mev$, $m_{a_0} \pm 20 \mev$, $m_{a_0} \pm 50 \mev$ and $m_{a_0} \pm 100 \mev$, respectively.

 In all the cases, we observe a peak corresponding to the TS.
 By looking at Fig.~\ref{Fig:Minv3}, we can see that integrating the double mass distribution over $M_{\rm inv}(\pi^- \eta)$ within the range $m_{a_0} \pm 100 \mev$ accounts for the whole strength of the $a_0(980)$ resonance,
 although one is introducing a bit of the $K\bar K$ in the final state apart from $\pi^- \eta$, as we discussed above referring to Fig.~\ref{Fig:tKK2}. 
 We interpret these results as indicative of what should be observed in the experiments.
 The shape of the TS is clearly observed. 

For the case where $M_{\rm inv}(\pi^- \eta) \in [m_{a_0}-100, m_{a_0}+100]\mev$, integrating over $M_{\rm inv}(\pi^+ a_0^-)$ in the range $[m_{\pi^+}+m_{a_0}, M_{J/\psi}-m_\phi]$ gives the branching ratio 
  \begin{equation}\label{eq:Br}
    {\rm Br}(J/\psi \to \phi \pi^+ a_0^- ) =1.07 \times 10^{-5},
  \end{equation}
to which we would associate an error of about $30\%$ from the uncertainties discussed in Sec.~\ref{subsec:IIA} when calculating the constant $C^2$ and the experimental error in the branching ratio of Eq.~\eqref{eq:bran} summing in quadrature. 
This estimate is realistic, as the only unknown magnitude required to evaluate the diagrams of Figs.~\ref{fig:mech2}(a) and \ref{fig:mech2}(c) is the $J/\psi \to \phi K^{*+} K^-$ amplitude, which we obtained from the experimental data.
This branching ratio is not small, given the copious production of $J/\psi$ at BESIII, which allows one to detect decays with branching fractions as small as $10^{-7}$ \cite{pdg}.
In view of this, we can only encourage the measurement of this reaction, which could also serve to clarify issues on the $J/\psi \to \pi^0 \eta \phi \,(\phi \to K \bar K)$ reaction \cite{besrecent} and its interpretation in Ref.~\cite{jingsakaiguo} in terms of a TS.
Actually, the $J/\psi \to \phi \pi^+ \pi^- \eta$ reaction has already been measured at BESIII \cite{besthis}.
However, the mass distributions that we propose here were not investigated.
Instead, the production modes of $\eta \phi f_0(980)$ and $\eta \phi f_1(1285)$ were investigated in that work.

Yet, we would like to call attention to a feature of the reaction of Ref.~\cite{{besthis}} of relevance to our work.
Indeed, in Fig.~5 of Ref.~\cite{{besthis}} there is a clear bump in the $\eta \pi^+ \pi^-$ mass distribution stretching from $1400\mev$ to $1530\mev$.
This bump was not unnoticed in Ref.~\cite{{besthis}} and was associated to the excitation of $\eta(1405)$, which has the $\eta \pi^+\pi^-$ as one of the decay modes.
In Table III of Ref.~\cite{{besthis}}, the branching ratio of the bump was estimated to be $(2.01\pm0.58\pm 0.82)\times 10^{-5}$.
It is interesting to see that twice our rate of Eq.~\eqref{eq:Br}, to account also for $\phi\pi^- a_0^+$ decay, with $30\%$ uncertainty, 
gives $(2.14\pm 0.64)\times 10^{-5}$, in perfect agreement with the strength of the experimental bump.
This coincidence, and the position of the peak compared to our Fig.~\ref{Fig:Minv3} give us strong arguments to encourage the reanalysis of this decay mode from the perspective given in the present work.
Let us recall that from a resonance formation perspective, $\eta \pi^+\pi^-$ cannot be $\eta \rho^0$, which would violate isospin conservation, but can be $a_0^- \pi^+$ (the mode studied here) and $a_0^+\pi^-$ which would have the same rate of production. 
The study of the $J/\psi \to \phi \pi^+ a_0^- \to \phi \pi^+ \pi^- \eta$ and $J/\psi \to \phi \pi^- a_0^+ \to \phi \pi^- \pi^+ \eta$ decay modes would allow one to make a comparison with the predictions made here and eventually, conclude the presence of the triangle singularity discussed in this work.

One should also recall that the $\eta(1405)\to \pi a_0 \to \pi\pi\eta$ and the isospin forbidden mode $\pi f_0(980)$ were also studied in Refs.~\cite{wuzou, fcaliang, wuwu, Acha1, Acha2} and shown to be dominated by a TS like the one discussed here, except that the $\eta(1405) \to K^* \bar K$ and the $J/\psi \to \phi K^* K$ vertices have different structures.
It should be thus possible to disentangle experimentally the $\eta(1405)$ excitation mode from the mechanism suggested here.
In this respect, works are already coming, providing methods to disentangle structures due to a TS or a resonance pole \cite{Co:2024bfl}.

\section{Conclusions}
We have conducted a study of the $J/\psi \to \phi \pi^+ a_0(980)^-\;(a_0^- \to \pi^- \eta)$ decay, showing that it develops a triangle singularity at $M_{\rm inv}(\pi^+ a_0^-)$ of about $1420 \mev$. 
The reaction proposed is motivated by the recent measurement at BESIII of the $J/\psi \to \eta \pi^0 \phi \; (\phi \to K \bar K)$ reaction \cite{besrecent}, that according to the work of Ref.~\cite{jingsakaiguo} also develops a triangle singularity, however, blurred by the tree level competing mechanisms and their interconnection due to the Coleman Norton theorem. 
In the reaction proposed, there is no tree level competing mechanism and then the TS appearing can be clearly interpreted. 
We evaluate the mass distributions in terms of $M_{\rm inv}(\pi^- \eta)$ and $M_{\rm inv}(\pi^+ a_0^-)$. In the $\pi^- \eta$ mass distribution we see a clear cusp structure, as observed in recent high statistics experiments, and in the $\pi^+ a_0^-$ mass distribution we observe the TS peak around $M_{\rm inv}(\pi^+ a_0^-)=1420 \mev$. By taking information for the needed $J/\psi \to \phi K^* \bar K$ amplitude from experiment, we are able to determine absolute rates for the reaction. 
Integrating the double mass distribution in the range of the $a_0(980)$ mass and in the range of the $\pi^+ a_0^-$ mass distribution, we predict a branching ratio for the reaction of the order of $10^{-5}$. 
Given the present rates of $J/\psi$ production at BESIII, where branching ratios of $10^{-7}$ can be measured, we advocate for the measurement of these mass distributions, that apart from showing a new example of a TS can also shed light on the interpretation of the recent BESIII measurements of the $J/\psi \to  \eta \pi^0 \phi \;(\phi \to K \bar K)$ reaction.
The realization of this task is more appealing since the $J/\psi \to \eta \pi^+ \pi^- \phi$ reaction was already studied at BESIII \cite{besthis}, although the decay modes discussed here were not addressed there.
We have discussed, however, that a peak seen in the $\eta \pi^+\pi^-$ mass distribution of this decay in the region $1400-1530 \mev$ is compatible with the signal that we have obtained from the TS, and encourage the experimental teams to look into the $\phi \pi^+ a_0^-$ and $\phi \pi^- a_0^+$ decay channels to further clarify this issue.

\begin{acknowledgments}
  J.M. Dias would like to express gratitude to Guangxi Normal University for the warm hospitality, as part of this work was conducted there.
  This work is partly supported by the National Natural Science Foundation of China under Grants No. 12365019, No. 11975083 and  No. 12175066, and by the Central Government Guidance Funds for Local Scientific and Technological Development, China (No. Guike ZY22096024).
  This work is also supported partly by the Natural Science Foundation of Changsha under Grant No. kq2208257 and the Natural Science Foundation of Hunan province under Grant No. 2023JJ30647 and the Natural Science Foundation of Guangxi province under Grant No. 2023JJA110076 (CWX).
  J.M. Dias acknowledges the support from the Chinese Academy of Sciences under Grant No. XDB34030000, and No. YSBR-101; by the National Key R\&D program of Chinese under Grant No. 2023YFA1606703; by NSFC under Grants No. 12125507, No. 12361141819, and No. 12047503. 
  This project has received funding from the European Union Horizon 2020 research and innovation programme under the program 2020-INFRAIA-2018-1, Grant Agreement No. 824093 of the STRONG-2020 project.
\end{acknowledgments}

\end{document}